\documentclass[traditabstract]{aa} 
\usepackage{graphicx}
\usepackage{natbib}
\bibpunct{(}{)}{;}{a}{}{,} 
\begin{document}
\title{Gran Telescopio Canarias OSIRIS Transiting Exoplanet Atmospheric Survey:  Detection
  of potassium in XO-2b from narrowband spectrophotometry\thanks{
Based on observations made with the Gran Telescopio Canarias (GTC),
installed in the Spanish Observatorio del Roque de los Muchachos of
the Instituto de Astrof\'\i sica de Canarias, in the island of La Palma,
and part of the large ESO program 182.C-2018.}
}

\author{D. K. Sing\inst{1}, J.-M. D{\'e}sert\inst{2}, 
                                J. J. Fortney\inst{3},
                                A. Lecavelier des Etangs\inst{4},
                                G. E. Ballester\inst{5}, 
                                J. Cepa\inst{6}, \\
                                D. Ehrenreich\inst{7}, 
                                M. L{\'o}pez-Morales\inst{8,9},
                                F. Pont\inst{1}, 
                                M. Shabram\inst{3}, \&
                                A. Vidal-Madjar\inst{4}
          }

\institute{
Astrophysics Group, School of Physics, University of Exeter, Stocker
  Road, Exeter, EX4 4QL\\email: sing@astro.ex.ac.uk 
\and
Harvard-Smithsonian Center for Astrophysics, 60 Garden Street,
Cambridge, MA 02138, USA 
\and
Department of Astronomy and Astrophysics, University of California,
Santa Cruz, CA 95064, USA 
\and
Institut d'Astrophysique de Paris, UMR7095 CNRS, Universite Pierre \&
Marie Curie, 98bis boulevard Arago, 75014 Paris, France 
\and
Lunar and Planetary Laboratory, University of Arizona, 
           Sonett Space Science Building, Tucson, AZ 85721-0063, USA 
\and
Departamento de Astrof\'\i sica, Universidad de La Laguna, Instituto
de Astrof\'\i sica de Canarias, La Laguna, Tenerife, Spain 
\and
Laboratoire d'Astrophysique de Grenoble, Universit\'e Joseph Fourier, CNRS (UMR 5571), BP 53, 38041 Grenoble cedex 9, France
\and
Carnegie Institution of Washington, Department of Terrestrial Magnetism, 5241 Broad Branch Road
NW, Washington, DC 20015, USA; Hubble Fellow
\and
Institut de Ciencies de l'Espai (CSIC-ICE), Campus UAB, Facultat de Ciencies, Torre C5, parell, 2a pl,
E-08193 Bellaterra, Barcelona, Spain
}

   \date{Received 13 August 2010/ Accepted 16 December 2010}

  \abstract{
  
We present Gran Telescopio Canarias (GTC) optical transit narrow-band
photometry of the hot-Jupiter exoplanet XO-2b using the OSIRIS
instrument.  This unique instrument has the capabilities to deliver high
cadence narrow-band photometric lightcurves, allowing us to probe
the atmospheric composition of hot Jupiters from the ground.  
The observations were taken during three transit events
which cover four wavelengths at spectral resolutions near 500, necessary for
observing atmospheric features, and have near-photon limited sub-mmag
precisions.  Precision narrow-band photometry on a large aperture
telescope allows for atmospheric transmission spectral features to be
observed for exoplanets around much fainter stars than those of the well studied targets
HD~209458b and HD~189733b, providing access to the majority of known
transiting planets.
For XO-2b, we measure planet-to-star radius contrasts of 
R$_{pl}$/R$_{\star}$=0.10508$\pm$0.00052  at 6792~\AA,
0.10640$\pm$0.00058 at 7582~\AA, and 0.10686$\pm$0.00060 at
7664.9~\AA, and 0.10362$\pm$0.00051 at 8839~\AA.
These measurements reveal significant spectral features at two
wavelengths, with an absorption level of 0.067$\pm$0.016\% at 7664.9 \AA\ due to atmospheric
potassium in the line core (a 4.1-$\sigma$ significance level), 
and an absorption level of 0.058$\pm$0.016\% at 7582 \AA, (a 3.6-$\sigma$
significance level).
When comparing our measurements to hot-Jupiter atmospheric models, we
find good agreement with models which are dominated in the optical by alkali metals.
This is the first evidence for potassium in an extrasolar
planet, an element that has long been theorized along with sodium to be a
dominant source of opacity at optical wavelengths for hot Jupiters.
}

\titlerunning{Detection of potassium in XO-2b from GTC spectrophotometry}
\authorrunning{Sing et. al}
\maketitle

%


\newpage
\section{Introduction}
Transiting hot-Jupiter exoplanets provide an excellent opportunity to 
detect and characterize exoplanetary atmospheres.
During a transit event, both the opaque body of the planet and its atmosphere
block light from the parent star, allowing the radius and its wavelength dependence to be
accurately measured, giving rise to the detection of atomic and molecular features in the atmosphere.  
In addition, secondary eclipse measurements can be used to obtain an exoplanet's
emission spectra, where such properties as the temperature, thermal structure,
and composition can be measured (e.g. \citealt{
2005Natur.434..740D, 
2007ApJ...658L.115G, 
2008ApJ...686.1341C, 
2009ApJ...707.1707R, 
2009A&A...493L..31S}) 
while orbital phase curves can probe the global temperature distribution \citep{
2007Natur.447..183K}. 

For highly irradiated planets, the atmosphere at optical wavelengths represents the
``window'' into the planet where the bulk of the stellar flux is deposited.  Thus, the
atmospheric properties at these wavelengths are directly linked to important global properties
such as atmospheric circulation, thermal inversion layers, and inflated planetary radii.  All
of these features may be linked within the hot-Jupiter exoplanetary
class \citep{
2008ApJ...678.1419F}, 
and transmission spectra can provide direct observational constraints and fundamental measurements of the atmospheric
properties at these crucial wavelengths.  In the optical, hot-Jupiter
transmission spectra are thought to be largely dominated 
by the opacity sources of the alkali metals sodium and potassium, with
both species predicted to be present since the initial model
calculations of highly irradiated gas giant exoplanets \citep{
2000ApJ...537..916S, 
2001ApJ...553.1006B, 
2001ApJ...560..413H}.  
Other strong but still unidentified optical absorbers could also play a role in the formation of stratospheres \citep{
2007ApJ...668L.171B, 
2008ApJ...673..526K}, 
with the leading candidates being TiO/VO and sulfur compounds \citep{
2003ApJ...594.1011H, 
2008ApJ...678.1419F, 
2008A&A...492..585D, 
2009ApJ...701L..20Z}. 
No candidate is yet a clear favorite to universally produce
hot-Jupiter stratospheres, both the TiO and sulfur explanations
have unresolved issues \citep{
2009ApJ...699.1487S} 
and stellar activity levels could also be important \citep{
2010ApJ...720.1569K}. 

To date, only two exoplanets have been ideal for optical atmospheric studies, HD~209458b and HD~189733b, 
mainly due to the stellar brightness (V$\sim7.8)$ along with the
large transit signals and low surface gravities.  Atmospheric sodium has been 
detected in HD209458b from data with {\it HST STIS} \citep{
2002ApJ...568..377C, 
2008ApJ...686..658S, 
2008ApJ...686..667S} 
and confirmed by ground-based spectrographs \citep{
2008A&A...487..357S}.  
In HD~189733b, a ground-based detection of sodium has also been made by \cite{
2008ApJ...673L..87R},  
with the wider overall optical spectrum consistent with a high altitude haze \citep{
2008MNRAS.385..109P, 
2009A&A...505..891S, 2009ApJ...699..478D}, 
thought to be due to Rayleigh scattering by small condensate
particles \citep{
2008A&A...481L..83L}. 
While sodium has been detected in both of these planets, the other
important alkali metal potassium has
not, though in principle both the ground-based and {{\it HST} spectra should have sufficient sensitivity.

Here we present the first results from GTC OSIRIS for the exoplanet XO-2b, part of our larger spectrophotometric optical
survey of transiting hot Jupiters.  
XO-2b is a 0.996$R_{Jup}$, 0.565 M$_{Jup}$ planet in a
2.6 day orbital period around a non-active $V$=11.2 early K dwarf star \citep{
2007ApJ...671.2115B}. 
XO-2b falls in the transition zone between the proposed ``pM/pL'' hot-Jupiter classification system of \cite{
2008ApJ...678.1419F}, 
and has been recently reported to perhaps have a 'weak' thermal inversion layer \citep{
2009ApJ...701..514M}. 
Our overall program
goals are to detect atmospheric features across a wide range of
hot-Jupiter atmospheres, enabling comparisons between the features
observed (comparative exoplanetology).
OSIRIS is an optical imager and spectrograph on the new 10.4 meter
GranTeCan (GTC) telescope, 
capable of extremely high-precision differential narrowband fast photometry, even for fainter transit hosting stars \citep{
2010arXiv1006.4599C}.  
In our program, transit events are observed with a narrowband tunable
filter, capable of simultaneously providing both the spectral resolution and
extremely high photometric precision necessary to detect atmospheric
transmission spectral features.
Multiple wavelengths are chosen to be maximally
sensitive to the expected hot-Jupiter atmospheric features, including sodium,
potassium, and titanium oxide.  
An important advantage to spectrophotometry is that the absolute transit
depths of each wavelength are measured, allowing for atmospheric features to be
compared together alongside existing broadband data at other
wavelengths.  In addition,
transit timing variation studies can also be made along with refining an exoplanet's system parameters.
In this paper, we describe the GTC XO-2b observations in \S 2, present the analysis of the transit
light curves in \S 3, discuss the results and compare to model atmospheres in \S 4, and conclude in \S 5.

\section{Observations}
We observed with the GTC 10.4 meter telescope installed in the Spanish Observatorio 
del Roque de los Muchachos of the Instituto de Astrof\'\i sica de Canarias on the island of La Palma, 
using the Optical System for Imaging and low Resolution Integrated Spectroscopy (OSIRIS) 
instrument during three separate nights in service mode.

\begin{figure}
 {\centering
  \includegraphics[width=0.49\textwidth]{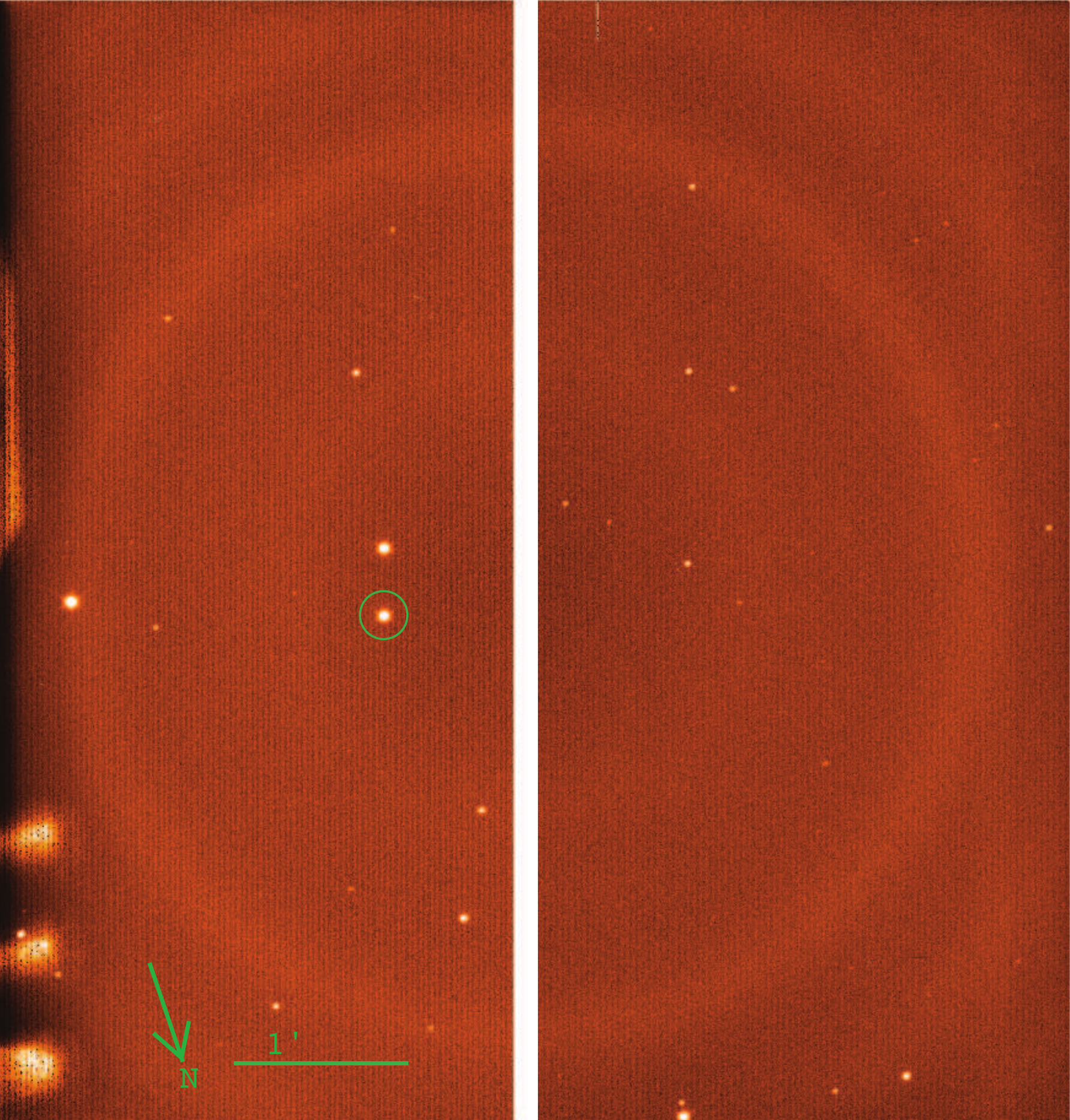}}
\caption[]{GTC Osiris CCD frame of XO-2 from 4 Dec 2009 taken with the
  tunable filter, with both CCD1 (left) and CCD2
  (right) shown.  XO-2 is the circled bright star located near the center of CCD1,
  with XO-2's stellar companion, used for our
  reference star, nearby and directly overhead.  The z-scale of the
  image has been
  set to best enhance the weak emission sky lines, which can
  be seen as ``rings'' around the image. Vignetting from components
  inside the instrument produce the artifacts seen at the very left edge of the image. }
\label{Figure:FOV}
\end{figure}

\subsection{GTC OSIRIS instrument setup}
The GTC OSIRIS instrument \citep{
2000SPIE.4008..623C, 
2003SPIE.4841.1739C} 
consists of a mosaic of two Marconi CCD detectors, 
each with 2048$\times$4096 pixels and a total
FOV of 7.8$'\times$8.5$'$, giving a plate scale of 0.127$''$/pixel.  The
CCD has a pixel well depth of $\sim$100,000 electrons and a 16 bit A/D
converter, which saturates at 65,536 counts.  For our observations, we
choose the 1$\times$1 binning mode with a full frame readout at a speed of
500kHz, which produces a readout overhead time of 31.15 seconds.
The readout speed of 500kHz has a gain of 1.46
e$^{-}$/ADU and a readout noise of 8 e$^{-}$.  While 500kHz is the 
noisiest readout speed (100 and 200kHz are available), it allows for
the most counts per image to be obtained due to a higher gain.  
Our initial service mode observations have been taken with the initially available full-frame readout mode, a
conservative observing approach for our initial observations, with
highly efficient windowing now becoming available.
In future observations, windowing (sub-arrays) will eliminate nearly all CCD readout time and provide better time-sampling with a near 100\% duty cycle.   

We used the tunable filter (TF), which allows for narrow-band imaging capable of specifying the
wavelength(s) of our target and reference stars.  The TF consists of a Fabry-Perot etalon,  
whose cavity separation can be adjusted, giving rise to a wide range of resolving powers, with OSIRIS capable of R=300 to 1000.
The unwanted orders from the etalon are then blocked by a set of broadband and mediumband (100-600 \AA\ FWHM) 
filters.  Different path-lengths through the optical system produce a wavelength dependance within the 
field of view, such that each image has ``rings''
of constant wavelength with respect to the TF-center, located near the
middle of the detector system (see Fig. \ref{Figure:FOV}).  To provide a maximum spectral resolution and
largest contrast for atmospheric signatures, we choose to observe with the
narrowest width available to all wavelengths of 12\AA.

For the XO-2b observations, we placed both the target and a chosen
nearby reference star, XO-2's stellar companion XO-2B, at the same radial distance
from the TF-center such that they would be observed at the same
wavelength (see Fig. \ref{Figure:FOV}).  
The companions XO-2 and XO-2B are separated by 31$''$ and cataloged as each having the same
spectral type (K0V) and have very similar magnitudes and colors, making XO-2B an
ideal reference star.
When searching for the strong atmospheric potassium line, 
successive images were taken alternating between two different wavelength positions,
carefully chosen to also avoid bright sky lines and strong telluric absorption lines, yet located close enough
together in wavelength to be within the same order-sorter filter.  As the tunable filter
can scan between wavelengths within a given order-sorter filter very
rapidly ($\sim$1 ms) and accurately ($\sim$0.1 \AA), we were able to
observe at two potassium line wavelengths during a single transit while avoiding the typical long overheads associated with switching between traditional filters. 

\begin{figure*}
 {\centering
  \includegraphics[width=0.8\textwidth]{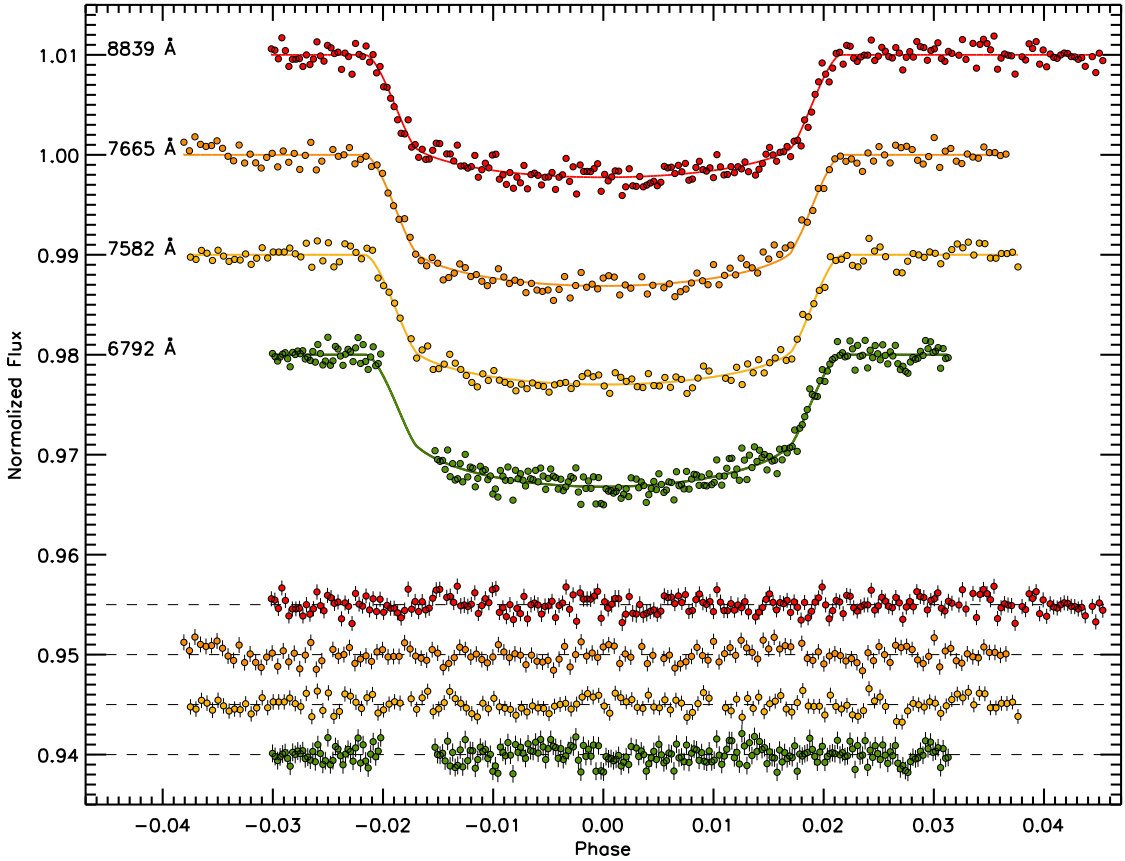}}
\caption[]{GTC OSIRIS narrow-band transit light curves at (top-to-bottom) 8839,
  7665, 7582, and 6792 \AA, with an arbitrary flux offset applied to each curve
  for clarity.  The best-fit models are also over-plotted,
  along with the residuals (bottom four curves) which have a y-axis
  error bar corresponding to the per-point flux uncertainty.  The
  light curves are near-Poisson limited, with per-point S/N levels between 1149 and 1351.}
\label{Figure:LC}
\end{figure*}

\subsection{Observing Log}
XO-2 was observed during the three transit events of 13 November 2009, 
04 December 2009, and 03 March 2010 (see Table \ref{TableObs}).

\begin{table} 
\centering
\caption{GTC OSIRIS XO-2 observations}
\label{TableObs}
\begin{tabular}{lllll}
\hline\hline  
 HJD start  &  $\lambda$ (\AA)  &  Number of  &     Integration time   \\ 
                 &                               &   exposures                 & (sec)  \\
\hline 
2455149.546        &  6792.0          & 249     &  20.47       \\ 
2455170.452        &  7664.9      & 147      &  25.47      \\ 
2455170.454         &   7582.0       &  149    &  25.47       \\ 
2455259.411         &   8839.0      &  223    &   40.67        \\ 
\hline
\end{tabular}
\end{table}

{\it 13 Nov. 2009}:  The seeing was very good during the observations, with values ranging from 0.6$''$ to 0.9$''$. 
The filter was tuned to a single wavelength of 6792.0 \AA, with 20.467 second exposure times used during the 
observing sequence with about 250 exposures acquired.
During the observations, there was an interruption of 20 min due to
problems with the M2 collimator, causing ingress to be largely lost.
The out of transit baseline flux was also truncated, compared to the
other nights, 
as time was lost with both the guiding setup and the TF entering a
forbidden range at the end of the night, where calibration becomes very unreliable.  However, we found 
there was still ample time to accurately measure the baseline flux both 
before and after the transit, and the interruptions and delays did not limit our analysis.
During the observations, the airmass decreased during the night from values of 1.678 to 1.08

{\it 4 Dec 2009}:  For the transit, XO-2 was observed at two TF wavelengths, switching successive images between 7664.89 and 7582 \AA.  
 The night was photometric, weather conditions very stable, and the
 seeing was very good during the sequence, 
with values ranging from 0.7$''$ to 0.9$''$.  Approximately 150 exposures at each
wavelength were acquired.  The total exposure time was 25.466
seconds for each image, producing successive frames at the same wavelength every 113.2
seconds, and an overall duty cycle of 45\%.
During the observations, the airmass decreased during the night from values of 2.158 to 1.081.

{\it 3 March 2010}:  Weather conditions were very stable, and the
seeing was good during the sequence, with values ranging from 0.7$''$
to 0.9$''$.   The filter was tuned to a single wavelength of 8839.0\AA, with 40.67 second exposure times used during the 
observing sequence with about 230 exposures acquired.
During the observations, the airmass increased during the night from values of 1.078 to 1.639.

For all three nights, exposure times were not modified during the sequence, maintaining the peak counts around 30000-35000 ADUs for both the reference and target stars.  When needed, defocusing was used to avoid
saturation in improved seeing conditions.   No recalibration of the TF was needed during the observing sequences, as the field 
rotator moved in a range where no variations are expected.  From data obtained during TF commissioning, 
possible variations in the central wavelength of the TF in the present observations are smaller than 1\AA, 
with real variations likely even lower.  In addition, image monitoring
every 10 min using strong sky lines show 
no variation in TF tuning during the observations.

For each night 100 bias and 100 well exposed dome flat-field images were taken ($\sim$30,000 ADU/pixel/image).  Our program uses
dome-flats, as opposed to sky-flats, as small scale pixel-to-pixel variations are a potentially important source of noise, requiring
a large number of well exposed flats to effectively remove it.  The
dome lights are not very homogeneous, in wavelength or spatially,
which complicates flat-fielding large-scale features, and an illumination correction was added to the final co-added flat-field image.  Our
observations were all specified to occur at the same pixel location (no
dithering), to help suppress flat-fielding errors, with the guiding
each night keeping the target's location within a few pixels.

\subsection{Reduction}

The bias frames and flat fields were combined using standard IRAF routines, and used to correct each image, before
aperture photometry was performed using IDL for the target and reference stars.  The
aperture location was determined by fitting a 2D Moffat function \citep{
1969A&A.....3..455M} 
to the point spread function (PSF) of 
the target and reference stars within each image.  The background sky was subtracted using the pixels from an annulus around each aperture.
After selecting a wide range of apertures and sky annulus ring sizes, aperture radius sizes between 18 and 20 pixels 
were found to be optimal for our observations, along with sky annulus sizes in the ranges of 70 pixels for the 
inner sky ring radius and 150 pixels for outer radius.  
For a given night, the same sized apertures were used for both the target and reference star.
From our observed PSFs, the theoretical maximum S/N aperture (which balances the increased 
readout noise of including more pixels with
 the decreasing number of photons away from the PSF center, see \citealt{2000hccd.book.....H}) was found to
 be 20 pixels, in good agreement with the empirically determined optimal aperture size.  
For the Dec 2009 observations at 7582 \AA, the average count level was observed to be 3.44$\times$10$^6$ counts
for the target XO-2 and 3.55$\times$10$^6$ counts for the reference
star XO-2B.  Including the errors from photon-noise, readout noise and sky subtraction, the average per-exposure uncertainty is found to be 
$\sigma_{XO2}$=0.00053 for XO-2 and $\sigma_{REF}$=0.00053 for the reference star, giving a 
differential photometric light-curve (LC) with an average per-exposure error of
$\sigma_{LC}$=0.00074, or a S/N per image of 1351.  The error
including photon, read noise, and sky background uncertainties were tabulated for each exposure, and used as the estimate of the
error in the differential photometric light-curve.  Similar values are
found for the other two nights, with an average S/N per image of 1167
for the 6792 \AA\ exposures and an average S/N of 1342 for the 8839
\AA\ exposures.\\
\indent The potassium line-core has nearby strong telluric O$_2$ absorption
lines, which could in principle affect the differential photometry.
However, we found that the resulting atmospheric differential extinction between the
target and reference star was negligible, $<4 \times 10^{-5}$, when
modeling the transmission function $t$
for each star through the atmosphere as a function of airmass, $t^{sec(z)}$
\citep{
2010A&A...523A..57V}. 

\begin{table} 
\centering
\caption{Three parameter non-linear limb darkening coefficients for
  GTC OSIRIS Tunable Filters}
\label{Table:LD}
\begin{tabular}{lllll}
\hline\hline  
 Wavelength (\AA)   &  ~~6792  &              ~~7582  &      ~7664.9 &   ~~8839  \\ 
\hline 
 $c_2$                             &  ~1.4055        &  ~1.4003   &  ~1.4036      &  ~1.4135     \\ 
 $c_3$                             & -1.0570       & -1.3042      &  -1.2957     &  -1.3454     \\ 
 $c_4$                             &  ~0.3453        &  ~0.4802    &  ~0.4745     &  ~0.4958    \\ 
\hline
\end{tabular}
\end{table}

\begin{table*} 
\caption{System parameters for XO-2b from individual light curve fits}
\label{Table:2}
\begin{centering}
\renewcommand{\footnoterule}{}  
\begin{tabular}{llllll}
\hline\hline  
Parameter  & Value\\
\hline 
central wavelength, $\lambda$ (\AA)                 & 6792.0                       &       7582.0                     &                7664.9           &                    8839.0  & \\
wavelength range, $\Delta\lambda$(\AA)          & 12                                &                  12                     &                       12              &              12         \\
planet/star radius contrast, $R$=R$_{pl}$/R$_{\star}$  &   0.1043$\pm$0.0021  &     0.1067$\pm$0.0015   &  0.1083$\pm$0.0017  &  0.1032$\pm$0.0012     \\ 
inclination, $i$ [deg]                                           & 88.4$\pm$1.8            & 87.23$\pm$0.99           &    86.09$\pm$0.79       &  88.5$\pm$1.9    &    \\ 
impact parameter, $b$=$a$cos$i$/R$_{\star}$    &  0.23$\pm$0.25         &      0.37$\pm$0.13        &  0.49$\pm$0.10             &   0.21$\pm$0.21 \\ 
system scale, $a/$R$_{\star}$                               &   8.14$\pm$0.43        &7.69$\pm$0.36              &  7.20$\pm$0.32             &8.05$\pm$0.35\\
stellar density, $\rho_{\star}$ [g cm$^{-3}$]          &   1.49$\pm$0.25        &     1.26$\pm$0.17        &     1.03$\pm$0.14         &     1.44$\pm$0.24  \\ 
linear limb-darkening coeff., $c_{2}$                   &   1.3943$\pm$0.048    &     1.412$\pm$0.053    &   1.491$\pm$0.061          &  1.448$\pm$0.043  \\ 
\hline
\end{tabular}
\end{centering}
\\
\end{table*}
\begin{table} 
\caption{Joint fit system parameters for XO-2b}
\label{Table:joint}
\begin{centering}
\renewcommand{\footnoterule}{}  
\begin{tabular}{ll}
\hline\hline  
Parameter & Value\\
\hline 
Period, $P^\dagger$ [days] &2.61586178$\pm$0.00000075\\
Mid-transit Time, $T(0)^\dagger$ [HJD] & 2454466.88457$\pm$0.00014\\
radius reference, $R_{ref}$=R$_{pl}$/R$_{\star}$        &   0.10362$\pm$0.00070  \\
altitude, $Z(\lambda=6792)$           &   0.00146$\pm$0.00075  \\
~~~~~~~~~~~~$Z(\lambda=7582)$  &   0.00278$\pm$0.00077  \\
~~~~~~~~~~~~$Z(\lambda=7664)$   &   0.00322$\pm$0.00079  \\
~~~~~~~~~~~~$Z(\lambda=8839)$   &   0 (fixed)  \\
inclination, $i$ [deg]                                               &    87.62$\pm$0.51    \\
system scale, $a$/R$_{\star}$                                   &   7.83$\pm$0.17   \\
impact parameter, $b$=$a$cos$i$/R$_{\star}$        &    0.324$\pm$0.070  \\
stellar density, $\rho_{\star}$ [g cm$^{-3}$]              &   1.328$\pm$0.088     \\
linear limb-darkening coeff., $c_{2}$                      &   fixed to Table \ref{Table:LD} values\\
$T(261)$ [HJD]  & 2455149.62450$\pm$0.00027  \\
$T(269)$ [HJD]  & 2455170.55140$\pm$0.00023  \\
$T(269)$ [HJD]  & 2455170.55181$\pm$0.00023  \\
$T(303)$ [HJD]  & 2455259.48978$\pm$0.00030 \\
\hline
\multicolumn{2}{c}{Using best fit values ($i$,  $a$/R$_{\star}$, $b$, \& $\rho_{\star}$)}\\
\hline
R$_{pl}$/R$_{\star}$($\lambda$)=$R$(6792\AA)   &   0.10508$\pm$0.00052  \\
~~~~~~~~~~~~~~~~$R$(7582\AA)         &   0.10640$\pm$0.00058  \\
~~~~~~~~~~~~~~~~$R$(7664\AA)         &   0.10686$\pm$0.00060  \\
~~~~~~~~~~~~~~~~$R$(8839\AA)         &   0.10362$\pm$0.00051  \\
\hline
\multicolumn{2}{c}{Previously determined model dependent parameters}\\
\hline
$R_{\star}$ [$\mathrm{R_{\odot}}$]    &  0.976$\pm$0.020$^{\ddag}$ \\
$R_{pl}$  [$\mathrm{R_{Jup}}$]  & 0.996$\pm$0.025$^{\ddag}$ \\
$M_{\star}$ [$\mathrm{M_{\odot}}$]    &  0.971$\pm$0.034$^{\ddag}$ \\
$M_{pl}$ [$\mathrm{M_{Jup}}$]   &  0.565$\pm$0.054$^{\ddag}$ \\
\hline
\end{tabular}
\end{centering}
\\
$^\dagger$fit along with the data from \cite{2007ApJ...671.2115B}  and \cite{2009AJ....137.4911F}\\  
$^{\ddag}$From \cite{2009AJ....137.4911F} \\
\end{table}

\section{Analysis}
\subsection{Transit light curve fits}
We modeled the transit light curve with the analytical transit
models of \cite{
2002ApJ...580L.171M},  
fitting for the central transit time, planet-to-star radius contrast,
limb-darkening, inclination, stellar density, and baseline flux. 
The errors on each datapoint were set to the tabulated values which
included photon noise, sky subtraction, and readout noise.
To account for the effects of limb-darkening on the transit light curve, we adopted the three parameter limb-darkening law, 
\begin{equation}  \frac{I(\mu)}{I(1)}=1 - c_2(1 - \mu) - c_3(1 - \mu^{3/2}) - c_4(1 - \mu^{2}) \end{equation}
calculating the coefficients  following \cite{
2010A&A...510A..21S} 
using the transmission through our narrow-band tunable filters (see table \ref{Table:LD}).  
We used the three parameter non-linear limb-darkening law, as it best
provides both the 
flexibility to capture the inherent non-linearity of the limb
darkening intensity profile and helps to avoid the 
deficient ATLAS model values near the limb (see \citealt{
2010A&A...510A..21S} 
for more details).  
This law has now successfully been used to model
high precision transit 
light curves including data from {\it HST/NICMOS} \citep{
2009A&A...505..891S}, 
{\it Spitzer} \citep{
2010A&A...516A..95H, Desert2010}, 
and CoRot.
In the fits, we also allowed the baseline flux level of each visit to vary in
time linearly, described by two fit parameters.  The linear trend
accounts for possible pixel-scale related drifts during the
observations, and we found that two 
nights exhibited a small baseline gradient.  The best fit parameters
were determined 
simultaneously with a Levenberg-Marquardt least-squares algorithm \citep{
2009ASPC..411..251M} 
using the unbinned data.
A few deviant points from each light curve were cut at the 3-$\sigma$ level in the residuals, likely due to cosmic rays.
The results from fitting each light curve separately are given in Table \ref{Table:2}.
In the separate light curve fits, we allowed the linear coefficient
term $c_2$ to vary freely, with the other two coefficients 
fixed to the predicted model values.  Choosing to fit only one
limb-darkening parameter ensures that the fits are both 
not significantly biased by adopting a stellar atmospheric model and
do not suffer from degeneracies between fitting 
multiple limb-darkening parameters.  
The fit values for the linear limb-darkening coefficients are measured
at the $\sim$4\% level, and are consistent within 1$\sigma$ of 
the model values (see Table \ref{Table:LD}), indicating that the Atlas models and three parameters law are performing well.  

We also performed a joint fit for the four light curves, simultaneously
fitting for the inclination $i$ and stellar density
$\rho_{\star}$, with the limb-darkening values fixed to the predicted
model values and the results plotted in Fig. \ref{Figure:LC}.  As
the precision and spectral resolution in our light-curves are
sufficient to observe atmospheric features, 
which cause measurable changes in the planet's radii, we jointly fit
each light curve with a common reference 
planetary radius, $R_{ref}$=R$_{pl}$/R$_{\star}$, as well an altitude
in units of stellar radius, $Z=Z(\lambda)$, above the fit reference radius value with 
\begin{equation}  R(\lambda)=Z(\lambda)+R_{ref}. \end{equation}
The transit light curve at 8839 \AA\ was used as the reference radius
value, setting the altitude $Z(8839)$=0, as it contained the smallest
measured radius.  The main advantage of 
jointly fitting $R_{ref}$ for the different color light curves, and
separately fitting $Z(\lambda)$, is that the uncertainties in 
fitting for the system parameters ($i$,  $a$/R$_{\star}$, $b$, \&
$\rho_{\star}$) do not affect the measurement 
(joint probability distribution) of $Z(\lambda)$, given that the
correlations between the system parameters and 
radius are largely wavelength independent.  We also quote the values
for the individual $R_{pl}/R_{\star}$ values for each light 
curve, when the system parameters are set to their best fit values.
The results of these joint 
fits are given in Table \ref{Table:joint}.

\subsection{Noise Estimation}
Each of our light curves achieve sub-mmag precision, with the RMS of the residuals between the best-fitting individual models and the data being
0.087\% at 6792 \AA\ (S/N of 1149),  
0.074\% at 7582 \AA\ (S/N of  1351),  
0.076\% at 7664 \AA\ (S/N of 1315), and 
0.083\% at 8839 \AA\ (S/N of 1204).  
These values compare very favorably to the theoretically estimated
error in \S 2, dominated by photon noise, indicating that our transit
light curves are regularly achieving between 90\% to 100\% of the
theoretical Poisson limit, without having to use 
de-correlation techniques to remove systematic errors.   
Other sources of noise are found to be negligible.  This includes scintillation,
as was also found to be the case with the large aperture of GTC by \cite{
2010arXiv1006.4599C}.  

We checked for the presence of systematic errors correlated in time (``red noise'', \citealt{
2006MNRAS.373..231P}) 
by checking that the binned residuals followed a $N^{-1/2}$ relation when
  binning in time by N points.  In the presence of red noise, the
  variance can be well modeled to follow a $\sigma^2=\sigma_w^2/N+\sigma_r^2$ relation,
  where $\sigma_w$ is the uncorrelated white noise component while
  $\sigma_r$ characterizes the red noise.  We found no significant
  evidence for red noise, when binning on timescales up to the ingress
  and egress duration, with all the determined $\sigma_r$ values
  below 1$\times10^{-4}$, a level corresponding to the limiting
  precision that our radii values can be measured within an individual
  light curve ($\sigma_{R^2}$$\sim$1.1$\times10^{-4}$).  
When possible, we also checked for the presence of red noise with the wavelet-based method of \cite{
2009ApJ...704...51C}, 
finding good general agreement with the binning technique.

\section{Discussion}
The system parameters determined with the GTC light curves are in
good agreement (at the $\sim$1-$\sigma$ level) with the most accurate
values for XO-2b previously published in the
literature, namely the z-band measurements from \cite{
2009AJ....137.4911F}. 
We choose our wavelengths to be maximally sensitive to the potassium
line, with the transits measured at 7664 and 7582 intended to measure
the K~I line core, while those at 6792 and 8834 \AA\ were chosen to
have minimum contributive opacities from K, Na, and H$_2$O.
Our joint-fit system parameters measure a significant altitude $z$ at two
wavelengths, with the measured potassium line center at 7664 \AA\ 
to be $z(7664)$= 0.00322$\pm$0.00079 $R_{\star}$, which is above the reference
altitude at the 4.1-$\sigma$
significance level (``absorption'' level of $\Delta (R_{pl}/R_{\star})^2=0.067\pm$0.016\%).
Assuming the stellar radius as given in Table~\ref{Table:joint}, the
potassium line core is seen to reach altitudes of 2190$\pm$540 km
above the 8839 \AA\ reference planetary radius.
A significant altitude is also measured at 7582~\AA, with $z(7582)$=0.00278$\pm$0.00077 $R_{\star}$ which
is above the reference zero altitude at the 3.6-$\sigma$ significance level.

The wavelength-dependent transit radius is defined as
the radius where the total slant optical depth reaches $\tau_{eq}$=0.56,
as determined in \cite{
2008A&A...481L..83L}, 
with the transmission spectra models of \cite{
2010ApJ...709.1396F} 
matching these results.  For an isothermal atmosphere, the effective altitude $z$ of the atmosphere as a
function of wavelength $\lambda$ is given in \cite{
2008A&A...481L..83L} to be
\begin{equation} z(\lambda)=H \mathrm{ln} \left(
    \frac{\varepsilon_{abs}  P_{ref} \sigma(\lambda)}{\tau_{eq}}
    \sqrt{\frac{2 \pi R_{pl}}{kT\mu g}} \right), \end{equation}
where 
$\varepsilon_{abs}$ is the abundance of dominating absorbing species, 
$P_{ref}$ is the pressure at the reference altitude,
$\sigma(\lambda)$ is the absorption cross section,
and $H$ is the atmospheric scale height.  The measured altitude parameter
$Z(\lambda)$ from \S3 relates to $z(\lambda)$ by $Z(\lambda)$=$z(\lambda)/R_{\star}$.

\begin{figure}
 {\centering
  \includegraphics[width=0.48\textwidth]{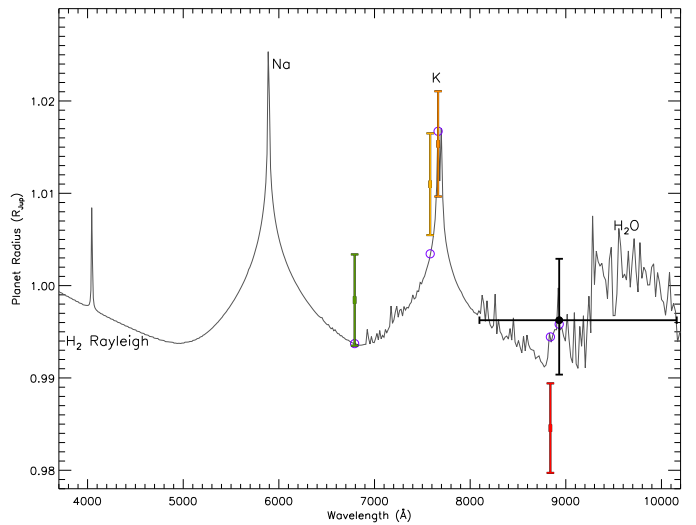}}
\caption[]{
XO-2b narrow-band GTC photometric results at four
  wavelengths (colored points) along with the z-band radius (black
  dot) of \cite{
2009AJ....137.4911F}. 
  The X-axis error bars indicate the wavelength range of the photometry,
  while the Y-axis error bars indicates the errors on the measured
  planetary radius, assuming a stellar radius of 0.976 R$_{\odot}$.
Also over-plotted is a 1500 K isothermal model (binned to R=500) of the transmission spectrum
from \cite{
2010ApJ...709.1396F} adapted for XO-2b, which is dominated by gaseous
Na and K.  TiO and VO opacities are removed in this model.
The open circles indicate the predicted filter-integrated model
values. }
\label{Fig1000}
\end{figure}

Assuming a temperature of
1500 K for XO-2, based on the broadband secondary eclipse measurements of \cite{
2009ApJ...701..514M}, 
and a surface gravity of 14.71 m/s$^2$, based on the planetary mass
and radius values found by \cite{
2009AJ....137.4911F}, 
we estimate an atmospheric scale height of $H$=370 km,
and $H/R_{\star}$=0.00054.  Thus, the individual precisions of our
wavelength dependent planetary radii measurements
($\sigma_{R_{pl}/R_{\star}}\sim$0.00055 \& $\sigma_{Z}\sim0.00075$) correspond to one
	  atmospheric scale height.  The atmospheric feature at 7664
          \AA\ thus spans 6$\pm$1.5
scale heights above the reference altitude.

\subsection{Atmospheric signatures of potassium}
We compare our best-fit planetary radius values to the recent
hot-Jupiter model atmospheric calculations of \cite{
2010ApJ...709.1396F}, 
along with XO-2b specifically generated models.  These models include a self-consistent treatment of radiative transfer and
chemical equilibrium of neutral species, though do
not include the formation of hazes.  Given our limited wavelength
coverage, such physically motivated models are necessary for
comparison with our data, as there is not enough data to justify freely fitting
for a large number of atmospheric parameters.  The comparisons here are
an effort to more broadly distinguish XO-2b's atmosphere between
different classes of models, given there are no prior constraints on
the basic composition or presence of clouds/hazes, as well as to compare
the specific predictions these models have for XO-2b.

The transmission spectrum calculations are performed for 1D
atmospheric pressure-temperature (P-T) profiles.  The model
atmospheres use the equilibrium chemical abundances, at solar
metallicity, described in \cite{
2002Icar..155..393L, 2006asup.book....1L}.  
The opacity database is described in \cite{
2008ApJS..174..504F}.   
The model atmosphere code has been used to generate P-T profiles for a
variety of close-in planets \citep{
2005ApJ...627L..69F,2006ApJ...642..495F,2008ApJ...678.1419F}.  
The transmission spectrum calculations are described in \cite{
2010ApJ...709.1396F}.  
For XO-2b, we find that the optical opacity is dominated by gaseous Na
and K, as is expected for many close-in planets over a broad
temperature range (e.g., \citealt{
2000ApJ...538..885S, 2000ApJ...537..916S}). 
These alkalis may be subjected to photo-ionization, but this is not
included in this study.

\begin{figure}
 {\centering
  \includegraphics[width=0.48\textwidth]{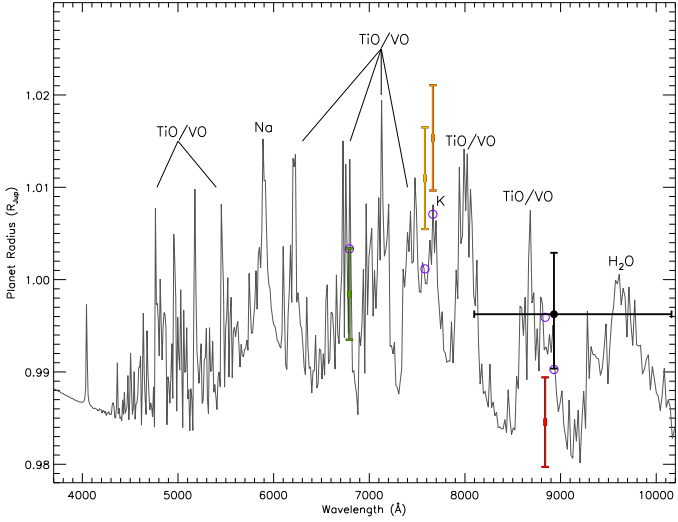}}
\caption[]{
Same as Fig. \ref{Fig1000}, but over-plotted with a 1500 K
isothermal model also containing TiO/VO opacity, producing many
strong absorption features.  The TiO/VO abundances in the model are from equilibrium
chemistry.  There is no evidence for TiO/VO in the GTC data, as the
model fits with TiO/VO are significantly worse than the
alkali-dominated models (see text).}
\label{Fig1500}
\end{figure}

The models are calculated wavelength-by-wavelength at a resolution of
R=3500, about seven times higher than the resolution of our data. 
For the comparisons to the GTC data, we assumed a value of $R_{\star}$=0.976 $M_{\odot}$ (Table 3), and
calculated the filter-integrated predictions from the high resolution model,
fitting only for an arbitrary model altitude shift.
Our GTC measurements are well matched with theoretical
models which include significant opacity from atmospheric potassium.  The
first indications to date that this species is present in hot-Jupiter
atmospheres.

{\it Isothermal models}: 
Comparing the observed atmospheric features to isothermal hydrostatic
uniform abundance models helps provide an overall understanding of the
features observed, and departures from those conditions.
Using equation 3, we adapted generic hot-Jupiter isothermal models
to values appropriate for XO-2.
We find an isothermal model dominated by alkali metals in the optical is
a substantially better fit than a grey ``flat''
atmosphere, as would
be expected if high-altitude clouds were dominant.  Adjusting only the
model reference altitude (i.e. the model normalization) in the fits to the GTC and z-band
points gives a $\chi^2$ of 22 with 4 degrees of freedom (DOF) for a
grey featureless atmosphere and a $\chi^2$ of
7.0 for a 1500 K isothermal alkali-dominated atmosphere (see
Fig. \ref{Fig1000}, with the model binned to R=500).  
With a smaller scale height, a cooler 1000 K isothermal model somewhat under-predicts the K~I core
levels, producing a $\chi^2$ fit of 9.3 with 4 DOF.

Models including TiO and VO in addition to K~I do not improve the
fits.  A 1500 K isothermal model including TiO and VO opacity at
chemical equilibrium abundances
gives a $\chi^2$ fit of 12.6 with 4 DOF (see Fig. \ref{Fig1500}).
In these models, TiO/VO effectively raise the altitudes of the 
non-potassium core model levels, degrading the fit.
These models still predict the 7664 \AA\ feature due to
potassium, as the low levels of TiO do not cover the potassium core
signature, with the K-doublet still visible in Fig. \ref{Fig1500}.  
Models completely dominated by only TiO/VO absorption, perhaps appropriate to
very hot-Jupiters such as Wasp-12, can be
definitively ruled out, as they provide a substantially poorer fit to the
data ($\chi^2$ of 32).
Sulfur compounds are unlikely to contribute at these longer
wavelengths.

{\it XO-2b specific models}:  We also compared our GTC planetary
radius values to models specifically calculated for the known
parameters of XO-2b, assuming chemical equilibrium at solar
metallicity.
A model using a dayside averaged temperature-pressure (T-P)
profile (see Fig. \ref{FigTP} and Fig.~\ref{FigFortney1000}) is our best fitting
model, giving a $\chi^2$ value of 6.6 for 4 DOF.
The cooler model transmission spectra using a planet-wide
averaged T-P profile (see Fig. \ref{FigTP}), performs moderately worse
giving a fit with a  $\chi^2$ of 8.0 for 4 DOF.  The planet-wide
averaged model predicts lower overall temperatures than the
day-side model, also resulting in a slight under-prediction of the
potassium line-core signature similar the 1000 K isothermal model.

\begin{figure}
 {\centering
  \includegraphics[width=0.48\textwidth]{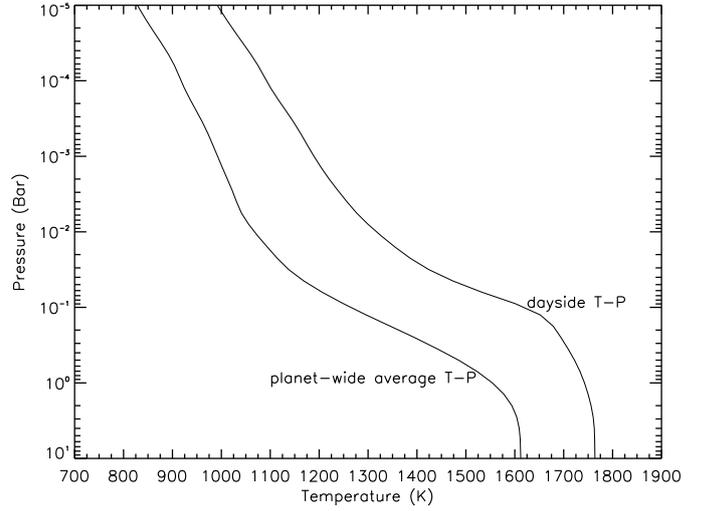}}
\caption[]{
Temperature-Pressure profiles for the XO-2b specific models, including a
planet-wide averaged profile (left) and a dayside averaged profile
(right).  The GTC transit data are sensitive to pressures from
$\sim$10$^{-1}$ bar at the lowest altitudes to $\sim$10$^{-3}$ bar at
  the highest altitudes in the potassium and sodium line-cores.
}
\label{FigTP}
\end{figure}

\begin{figure*}
\begin{center}
  \includegraphics[width=0.8\textwidth]{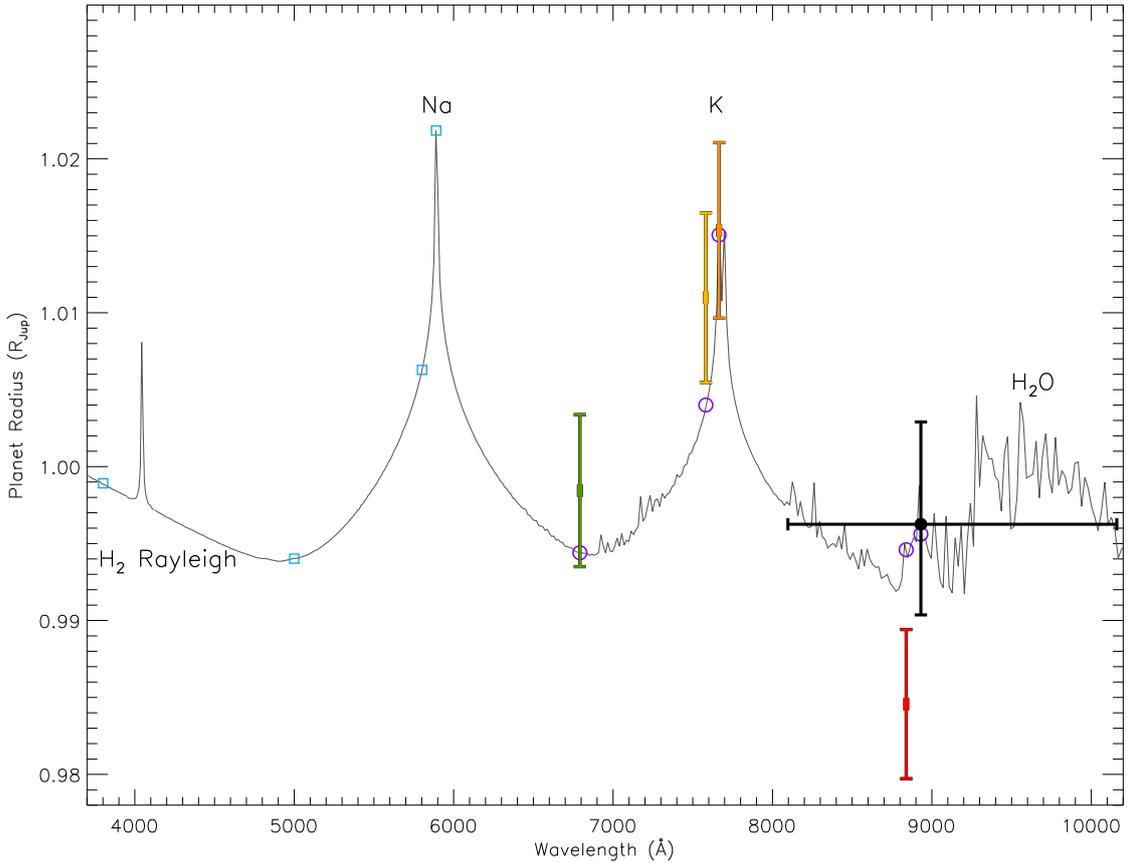}
\end{center}
\caption[]{Same as Fig. \ref{Fig1000} but over-plotted with a XO-2b specific model of the transmission spectrum, which
  uses a dayside-average temperature-pressure profile.  The model is dominated in the
  optical by Na and K absorption signatures.  Compared to the isothermal model of Fig. \ref{Fig1000}, the lower temperatures at higher-altitudes in
this model lead to a better fit of the K-core 7664.9 \AA\ measurement.
Further observations are planned at blue wavelengths (blue boxes),
  which will be highly sensitive to atmospheric sodium, and will help provide a
  more complete picture of the transmission spectrum.}
\label{FigFortney1000}
\end{figure*}
Both XO-2b specific models and the non-TiO isothermal models predict that the opacity from potassium,
sodium, and H$_2$ Rayleigh scattering dominate the optical, with H$_2$O absorption
features prominent red-ward of $\sim$9000 \AA.  
From our data, the observed strong potassium opacity in the optical and lack of
significant TiO is in general
agreement with XO-2b being classified as a transition ``pM/pL''
hot-Jupiter.
While the GTC data does not provide particularly strong temperature constraints, we
find that temperatures in the range of $\sim$1500 K from the isothermal and
day-side models offer slightly better fits than the cooler planet-wide
and 1000 K models.
Using 1500 K temperatures or a specific T-P profile from the dayside would at first seem at odds with transit
spectral data, associated with the planetary limb.  However, 3D hot-Jupiter models have recently shown
that the warmer dayside temperatures can increase the atmospheric scale height and
effectively ``puff up''  the dayside atmosphere, covering 
both the cooler planetary limb and night-side transmission spectral signatures \citep{
2010ApJ...709.1396F}.
Our optical transit data are sensitive to pressures from
$\sim$10$^{-1}$ bar at the lowest altitudes between the Na~I and K~I line
wings down to $\sim$10$^{-3}$ bar at
the highest altitudes in the potassium and sodium line-cores (assuming
solar abundances).  This
is roughly the same pressure range that Spitzer secondary-eclipse
emission spectra measurements are sensitive to, given that the smaller
cross section in the optical (compared to the IR) is compensated by
the larger horizontal integrated density of the slant-transit
geometry.  These similarities have the potential to help in identifying unknown
absorbers thought to be responsible for day-side inversions, as their 
signatures could appear in optical transmission spectra.
\cite{
2009ApJ...701..514M} 
reported a ``weak'' temperature inversion layer for XO-2b, with 
models placing it near a 10$^{-2}$ bar level.  Our models lack
thermal inversions, though  
we do find similar $\sim$1500 K temperatures as models containing a
thermal inversion \citep{
2009ApJ...701..514M}.   
 Given that non-inverted
models still fit the Spitzer IRAC photometry at 2.2-$\sigma$ levels
or better, and the average limb profile may still be sufficiently different
than the day-side and lack detectable inversions, we do not
consider the difference as significant.
Further secondary eclipse measurements will be necessary to definitively determine the
presence or absence of a thermal inversion layer in XO-2b.

\subsection{Limits on stellar activity}
A transmission spectrum in active stars can be affected from the
presence of stellar spots, both
when a spot is occulted by the planet and by the presence of non-occulted
spots during the epoch of the transit observations \citep{
2008MNRAS.385..109P, 
2010ApJ...721.1861A}. 
A non-occulted stellar spot changes the baseline stellar flux with which the
transit is normalized, such that different epochs with different amounts
of spots will show small but non-negligible radius changes.
For the XO-2b observations, sufficiently large stellar activity could in principle affect the
measured radii at the reference wavelengths (6792 \& 8839 \AA) compared to the potassium
line core wavelengths, as they were not taken simultaneously during a 
transit event. 
However, in the absence of strong stellar activity, multiple transits spanning long
epochs (i.e. non-simultaneous transits) can still be used and fit together at
high precision, without the need for a stellar activity related
correction.  

\cite{
2009A&A...505.1277C} 
and \cite{
Desert2010}  
proposed that the impact of rotation modulated spots on the final
fitted radius could be estimated following
\begin{equation}  \frac{\Delta
    R_{pl}/R_{\star}}{(R_{pl}/R_{\star})_{True}}= \frac{\alpha f_{\lambda}}{2},  \end{equation}
where $f_{\lambda}$ is the relative variation of the stellar flux due
to stellar activity (i.e. rotation modulated spots) and $\alpha$ is an
empirically determined constant of order unity.  In the hypothesis that
the stellar surface brightness outside dark stellar spots are not
modified by the activity level, the parameter $\alpha$ is given by $-1$.
The range of $f_{\lambda}$ between active and non-active stars is very large,
spanning many orders of magnitude.  For the highly active star
HD~189733, $f_{\lambda}\sim2-3\%$ in the visible \citep{
2008AJ....135...68H}, 
while for the non-active star TrES-2, $f_{\lambda}$ was found to have
an amazingly low value of (5.18$\pm$0.19)$\times10^{-6}$ from Kepler data \citep{ 
2010arXiv1006.5680K}. 
\cite{2010arXiv1009.1840C} 
studied the stellar photometric variability for stars in the Kepler
sample, finding a bi-model distribution between active an non-active
stars.  K-dwarfs with low activity (2/3 of the sample) were found to typically display
a stochastic photometric variability of less than 1 mmag, while active
stars were often found to be periodic on timescales of days to weeks.
For the HST transit observations of the highly active star HD~189733 \citep{
2008MNRAS.385..109P, 
2009A&A...505..891S, Desert2010}, 
an observational strategy was adopted to counter this effect, which
involved monitoring the stellar
activity level.  In this way, the stellar baseline flux was known at every
epoch of the transit observations, and could then be used to correctly
normalize the data.  For non-active stars such as TrES-2 or XO-2, this strategy
may not be needed as the transit-fit planetary radii values will not
significantly change from one epoch to another. 

With the TrES-2b Kepler data, and such a low level of activity, eighteen optical transits were
fit together by \cite{
2010arXiv1006.5680K},  
who found no signs of epoch-dependent radius changes, as would
be apparent from strong activity-related effects.  The star TrES-2 has very
similar activity indices as XO-2, with \cite{
2010ApJ...720.1569K} 
measuring the stellar activity indices of the Ca~II H \& K line
strength to be log($R'_{HK}$)=$-$4.988 for XO-2 and $-$4.949 for TrES-2, while the 
S-value of XO-2 is 0.173 and 0.165 for TrES-2.  Similar activity levels as TrES-2 suggests 
activity-related radii changes in XO-2 are also negligible.
The final TrES-2 combined transit light curve was successfully fit down to the
20 ppm level, which is nearly an
order of magnitude below our required levels for XO-2b ($\sim$100 ppm).
Hat-P7 and Kepler-4,5,6,7,8 also show a low 
photometric variability around the $\sim$0.2 mmag level, with
  the transits removed when calculating the light curve statistics \citep{
2010arXiv1009.1840C},  
and all have similar activity indices as XO-2 \citep{
2010ApJ...720.1569K}. 

For XO-2, there is also no obvious evidence for occulted spots in any of
our sub-mmag precision light curves 
consistent with a non-active star, though non-occulted spot features could be clustered at different latitudes.
In addition, no spot-induced rotational modulations have yet been reported, with
the discovery XO-2 light-curves seemingly constant down to a noise-limited $\sim$7 mmag
level \citep{
2007ApJ...671.2115B}. 
Assuming $f_{\lambda}<0.007$ for XO-2 and $\alpha=-1$, we find
$\Delta(R_{pl}/R_{\star})<0.00036$, which is below the limiting precision
level of our measurements, and below the estimated size of one atmospheric scale height.
While current evidence points towards XO-2 as being a non-active
  star with a likely small photometric variability, further activity
  monitoring of this and other transit hosting stars can play an
  important role in assessing the true variability.

\subsection{Transit Timing Studies}
We used the four central transit times of our GTC data along with the
transit times of \cite{
2007ApJ...671.2115B} 
and \cite{
2009AJ....137.4911F}, 
to determine the transit ephemeris, using a linear function of the
period $P$ and transit epoch $E$,
\begin{equation} T(E) = T(0) + EP. \end{equation}
The results are given in Table \ref{Table:joint}, where we find a
period of $P = 2.61586178\pm0.00000075$ (days)
and a mid-transit time of $T(0)=2454466.88457\pm$0.00014 (HJD).
A fit with a linear ephemeris is in general an unsatisfactory fit to the data,
having a $\chi^2$ of 42 for 20 degrees of freedom ($\chi^2_{\nu}$=2.1).
The fit period is within 1-$\sigma$ of the \cite{
2007ApJ...671.2115B} 
value of 2.615857$\pm$0.000005 days, but 
3-$\sigma$ away from the period of 2.615819$\pm$0.000014 days \cite{
2009AJ....137.4911F} 
found when fitting for the ephemeris with just their six transit times.
Similar to the conclusions of \cite{
2009AJ....137.4911F}, 
we find it is not clear whether to interpret the poor linear fit and
differences in period to a genuine period variation from the 
presence of a second planet, or whether the uncertainties in all the
timings are underestimated by $\sim\sqrt{2}$.  Given that the
observed variations are so close to the per-transit noise level, an unambiguous
transit timing variation and detection of a second planet free of
systematic errors will require more observations with generally improved
timing errors.  GTC OSIRIS observations in sub-array mode should
help provide the necessary precision, as the duty cycle can be increased to near 100\%.

\section{Conclusions}
We present GTC OSIRIS observations for three transits of XO-2b, at four carefully
selected narrowband wavelengths, which have revealed significant absorption
by atmospheric potassium.  
This is the first evidence for this important alkali
metal in an extrasolar planet, which has long been theorized along with sodium to be a
dominant source of opacity at optical wavelengths.  
Identification of atmospheric contents is one of the first steps to
understanding the nature of exoplanetary atmospheres.  The presence (or lack) of
key species can give clues on the exoplanet's temperature, presence of
condensate clouds, chemistry, and can potentially be used to type different atmospheres
and place exoplanets within different sub-classes.
Clear gaseous hot-Jupiter
atmospheres are thought to have low geometric albedos, appearing very
dark at visible wavelengths, with broad pressure broadened potassium and sodium
line-wings efficiently absorbing the incoming stellar radiation \citep{
2000ApJ...538..885S, 2000ApJ...540..504S}.  
A detection of potassium in XO-2b helps support this overall picture.
Potassium and sodium could also play important roles in inflating hot
Jupiters through ohmic dissipation mechanisms \citep{
2010ApJ...714L.238B}. 
Notably, there is also no particular evidence for K~I photoionization,
which could have been evident by depletion at high altitudes.

Narrow-band spectrophotometry is a robust method
for obtaining precision differential photometric transit light-curves which have a 
spectral resolution sufficient to detect atmospheric transmission spectral features.   
OSIRIS is the first instrument capable of producing sub-mmag
high-speed spectrophotometry, and coupled with the large 10.4 meter
aperture of GTC, is capable of detecting spectral features in transmission spectra
on exoplanets such as XO-2b, whose host star is 3.5 magnitudes fainter
in the optical than those of the
well studied transiting hot Jupiters HD~189733b and HD~209458b.
XO-2b now joins those two exoplanets as the only hot-Jupiters where direct constraints on
the atmospheric opacity in the optical can be made.  
Detecting and studying the atmospheres for such fainter targets will be
critical for a broader understanding of the hot-Jupiter class as a
whole, as most of the known transiting planets have been found at magnitudes $>$10.
Further observations from our program, including planned XO-2b measurements at blue wavelengths, can
help conclude whether the optical atmospheres of these exoplanets are dominated by
the two alkali metals sodium and potassium together, search for
Rayleigh scattering, and place limits on
other optical absorbers thought to be linked to thermal inversion layers.


\begin{acknowledgements}
We thank the entire GTC staff and especially Antonio Cabrera Lavers for his continued excellent work
executing this program.
D.K.S. is pleased to note that part of this work was performed while
in residence at the Kavli Institute for Theoretical Physics, funded by the NSF through grant No. PHY05-51164.
This work was partially supported by the Spanish Plan Nacional de
Astronom\'\i a y Astrof\'\i sica under grant AYA2008-06311-C02-01.
D.E. is supported by the Centre National d'\'Etudes Spatiales (CNES).
M.L.M. acknowledges support from NASA through Hubble Fellowship grant HF-01210.01-A/HF-51233.01
awarded by the STScI, which is operated by the AURA, Inc. for NASA, under contract
NAS5-26555.
\end{acknowledgements}

{\it Note Added}.  Shortly before submission, we became aware of independent, but
similar observations of another exoplanet \citep{
2010Colon}.
Future observations
of these and additional exoplanets will enable comparisons of the
atmospheric composition and structure, as well as studies of potential
correlations with other planet or host star properties.

\bibliographystyle{aa} 
\bibliography{Sing.GTC.XO2} 

\end{document}